\documentclass[reprint,aps,prb,superscriptaddress]{revtex4-1}
\usepackage{graphicx}
\usepackage[version=4]{mhchem}
\usepackage[colorlinks, citecolor=blue, linkcolor=blue, urlcolor=blue, breaklinks]{hyperref}
\usepackage{siunitx, xspace}
\DeclareSIUnit[]\muB{\text{\ensuremath{\mu_{\textup{B}}}}}
\bibpunct{[}{]}{;}{n}{}{}

\begin{document}

\title{Importance of effective dimensionality in manganese pnictides}

\author{Manuel Zingl}
\email[]{manuel.zingl@tugraz.at}
\affiliation{Institute of Theoretical and Computational Physics,
Graz University of Technology, NAWI Graz, 8010 Graz, Austria}
\author{Elias Assmann}
\affiliation{Institute of Theoretical and Computational Physics,
Graz University of Technology, NAWI Graz, 8010 Graz, Austria}
\author{Priyanka Seth}
\affiliation{Institut de Physique Th\'eorique (IPhT), CEA, CNRS,
UMR CNRS 3681, 91191 Gif-sur-Yvette, France}
\author{Igor Krivenko}
\affiliation{Institut f{\"u}r Theoretische Physik, Uni. Hamburg, Jungiusstra{\ss}e 9, 20355 Hamburg, Germany}
\author{Markus Aichhorn}
\affiliation{Institute of Theoretical and Computational Physics,
Graz University of Technology, NAWI Graz, 8010 Graz, Austria}

\date{\today}

\begin{abstract}
In this paper we investigate the two manganese pnictides \ce{BaMn2As2}
and \ce{LaMnAsO}, using fully charge self-consistent density functional 
plus dynamical mean-field theory calculations. These systems have a nominally 
half-filled $d$ shell, and as a consequence, electronic correlations are strong, 
placing these compounds at the verge of a metal-insulator transition. Although 
their crystal structure is composed of similar building blocks, our analysis 
shows that the two materials exhibit a very different effective dimensionality, 
\ce{LaMnAsO} being a quasi-two-dimensional material in contrast to the much more 
three-dimensional \ce{BaMn2As2}. We demonstrate that the experimentally observed 
differences in the N\'eel temperature, the band gap, and the optical properties 
of the manganese compounds under consideration can be traced back to exactly this 
effective dimensionality. Our calculations show excellent agreement with measured 
optical spectra.
\end{abstract}


\maketitle

\section{\label{sec:Intr}Introduction}
The Mott phenomenon, the occurrence of an insulating state solely due to 
electronic correlations, is among the most intensively studied effects in correlated 
solid state physics~\cite{imadarmp}. This insulating state can occur in situations where 
simple band theory would not allow it, e.g., for an odd number of electrons per unit cell 
in the absence of symmetry breaking. It is interesting not only in its own right but also 
as it is the host for other fascinating phenomena, the best-known example being high
temperature superconductivity in cuprate oxides. There, injecting charge carriers into this 
Mott insulating state by chemical doping creates a non-Fermi-liquid state which becomes 
superconducting at low temperatures~\cite{RevModPhys.78.17}.

In recent years another class of high-temperature superconductors, the iron-based pnictide 
and chalcogenide materials, has been identified~\cite{kamihara1}. In contrast to the cuprates, 
they are intrinsically multiband systems with the whole $3d$ manifold being relevant for the
electronic properties~\cite{haule1,LaFe01}. These iron-based materials share common building 
blocks, the iron-pnictogen or iron-chalcogen layers, and a nominal electronic configuration of 
six electrons in the five Fe-$3d$ bands, which places them in the ``Hund's metal'' 
regime~\cite{haule2009,georges2013_hund,Janus}. As a result of this band filling, these materials 
have very low coherence scales and sizable correlations, without, however, being close to a Mott 
metal-insulator transition.

Since superconductivity arises from quantum fluctuations in the normal state, we must understand 
the physical properties of the relevant parent compounds before we can hope to understand 
superconductivity. In this paper we therefore investigate the two \emph{manganese} pnictide
compounds \ce{BaMn2As2} and \ce{LaMnAsO}, which are isostructural to the iron-based superconductors 
\ce{BaFe2As2} and \ce{LaFeAsO}, but host only five electrons in the five Mn-$3d$ bands. These systems 
can be seen as the pnictide analog of the undoped parent compounds of the cuprate high-temperature 
superconductors~\cite{IshidaLiebsch, demedici_selective2014}. The half-filled Mn-$3d$ shells of these 
compounds promote Mott physics~\cite{Ba14,LaMnPO01,Ba07}. Efforts to induce metallicity by 
pressure~\cite{Ba08} or doping~\cite{Ba04, Ba05, Ba11, Ba12, Ba13, Ba15, La03} have been to some 
extent successful. While no superconducting state has been demonstrated conclusively~\cite{Ba08}, 
the manganese pnictides still feature fascinating properties such as giant magnetoresistance~\cite{La08,La06},
large Seebeck coefficients~\cite{La04, La02, Ba03, Ba09}, and strongly enhanced magnetism with 
antiferromagnetic (AFM) order persisting up to elevated temperatures~\cite{La06, Ba06}. In contrast to 
the related iron pnictides, both manganese pnictides investigated here are semiconductors, as shown in 
optical and conductivity measurements~\cite{La01,La02,Ba01, Ba02, Ba03, Ba08, Ba09, Ba11, Ba14}. 

When considering an insulating state in a half-filled system the question arises whether this state 
occurs due to electronic correlations alone (Mott mechanism) or because of symmetry breaking such as 
magnetism (Slater mechanism). In many correlated materials both mechanisms are at work and sometimes 
are of similar importance. Furthermore, it has been shown that the proximity to a Mott transition can strongly 
increase the magnetic ordering temperature~\cite{Neel04}. For instance, 1111 manganese pnictides including 
\ce{LaMnAsO} ($T_{\text{N}}$ $\approx$ \SI{350}{\kelvin}~\cite{La08, La04, La10}) and some 122 manganese pnictides, like 
\ce{BaMn2P2} ($T_{\text{N}} > \SI{750}{\kelvin}$~\cite{BaMn2P201}) and \ce{BaMn2As2} ($T_{\text{N}} = \SI{625}{\kelvin}$~\cite{Ba06}), 
remain ordered well above room temperature. N\'eel temperatures of this order naturally call for a closer 
investigation of the underlying mechanisms.

\begin{figure}[t]
\centering
\includegraphics[width=0.85\linewidth]{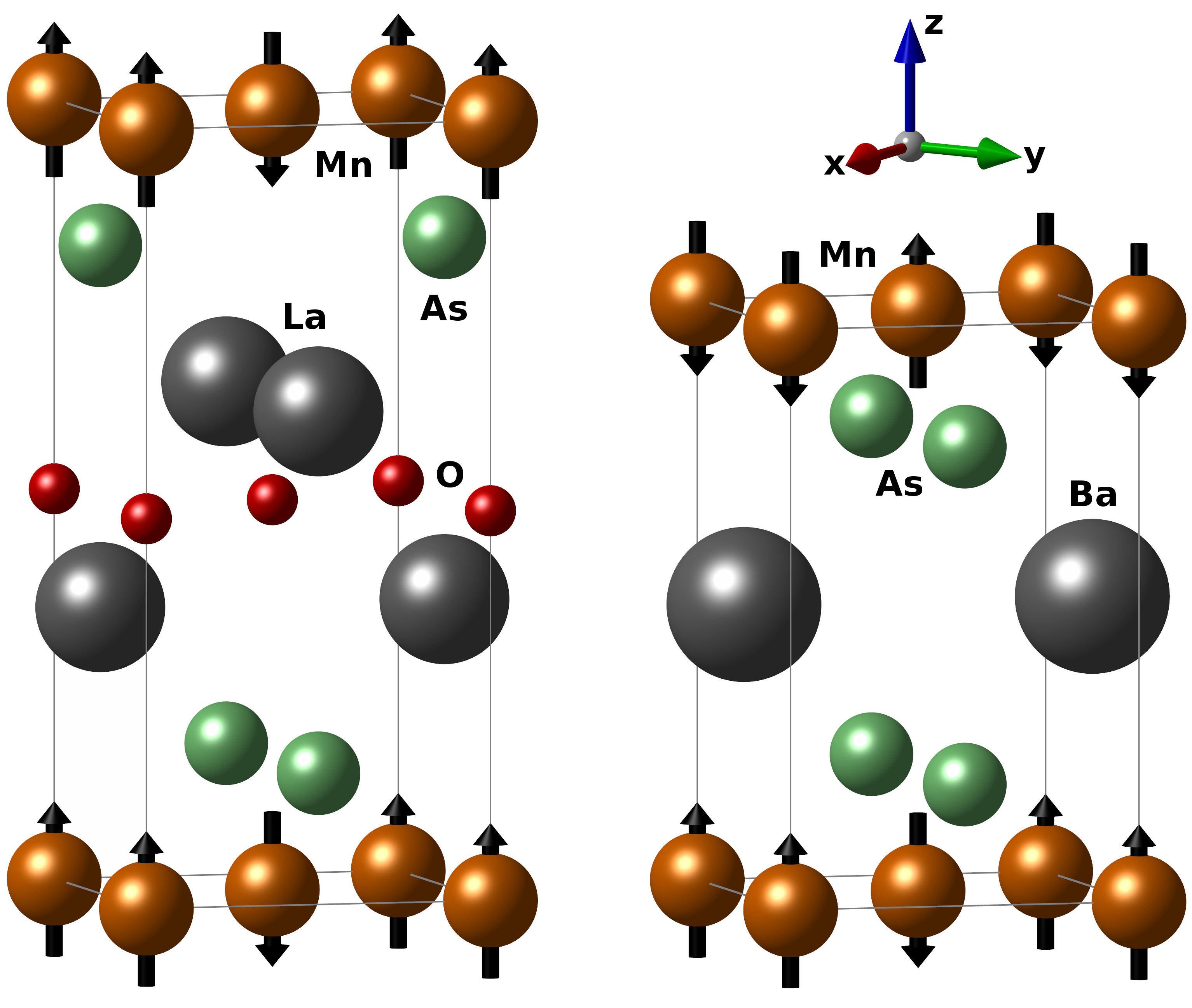}
\caption{\label{fig:crysstruct}Crystal and magnetic structure of \ce{LaMnAsO} (left) and \ce{BaMn2As2} (right) 
drawn with VESTA~\cite{VESTA}. The black arrows represent the Mn spins in the antiferromagnetic states of 
\ce{LaMnAsO}~\cite{La06} (C-type: ferromagnetically stacked antiferromagnetic planes) and \ce{BaMn2As2}~\cite{Ba06} 
(G-type: alternating in all directions). We choose a coordinate system where the $x$ and $y$ axes point towards 
the nearest-neighbor Mn atoms.}
\end{figure}

We will show that \ce{LaMnAsO} and \ce{BaMn2As2} are both close to a metal-insulator transition. However, 
there are differences in the effective dimensionality of the two compounds, which will turn out to be 
decisive for their properties. Specifically, \ce{BaMn2As2} crystallizes in a \ce{ThCr2Si2}-type structure 
and shows G-type AFM (antiferromagnetic in all directions, see Fig.~\ref{fig:crysstruct} right) and a large 
magnetic moment of \SI{3.9}{\muB/Mn}~\cite{Ba06}. \ce{LaMnAsO}, with its \ce{ZrCuSiAs} structure, features 
antiferromagnetic Mn~planes with a magnetic moment of \SI{3.6}{\muB/Mn}~\cite{La06,La06_sup}, but the coupling 
between planes is ferromagnetic (C-type AFM, shown in Fig.~\ref{fig:crysstruct} left). Although both compounds 
share \mbox{Mn-As} layers with comparable Mn-Mn distances, the different layer stacking and the larger Mn 
interlayer spacing turn \ce{LaMnAsO} into a quasi-two-dimensional compound~\cite{La05, La09, La10}, while 
\ce{BaMn2As2} is much more three dimensional~\cite{Ba03,Ba10}. A dependence of the physical properties on the 
effective dimensionality has also been observed, e.g., in the iron pnictides~\cite{andersenboeri, dimironpnic}.

This paper is organized as follows. Section~\ref{sec:Meth} is dedicated to an outline of the methods and parameters we 
use. In Sec.~\ref{sec:Resu}, our results on the electronic structure, magnetic, and optical properties are presented 
and compared to experimental values where available. Finally, we conclude in Sec.~\ref{sec:Conc}.

\section{\label{sec:Meth}Methods}
DFT+DMFT~\cite{DFTDMFT1}, the combination of density functional theory (DFT) 
and dynamical mean-field theory (DMFT)~\cite{DMFT0, DMFT1, DMFT2, DMFT3}, 
is used as a theoretical framework for the electronic structure calculations 
presented in this paper. Unless otherwise stated, calculations were carried 
out with the fully charge self-consistent implementation of the TRIQS/DFTTools 
package (v1.3)~\cite{TRIQS/DFTTools,LaFe01,TRIQS/DFTTools2}, which is based on 
Wien2k (v14.2)~\cite{Wien2k1} and the TRIQS library (v1.3)~\cite{TRIQS}. 
We use crystal structures measured at \SI{300}{\kelvin} for \ce{BaMn2As2}~\cite{Ba06} 
and \SI{290}{\kelvin} for \ce{LaMnAsO}~\cite{La08}. The same crystal structures are 
used for antiferromagnetic and paramagnetic calculations as no structural phase 
transition accompanies the magnetic transition in either compound~\cite{La06, Ba06}.  
For the antiferromagnetic calculations we use the experimentally determined magnetic 
orderings (see Fig.~\ref{fig:crysstruct}), which are also predicted by total-energy DFT 
calculations~\cite{Ba03,La05}.  Note that due to the G-type ordering the magnetic unit 
cell is doubled in the $z$ direction in \ce{BaMn2As2}. For the DFT part of the fully 
charge self-consistent calculations we use \num{10000} $k$ points in the full Brillouin 
zone (BZ) and employ the standard Perdew-Burke-Ernzerhof (PBE)~\cite{PBE} generalized 
gradient approximation (GGA) for the exchange-correlation functional.

From the DFT Bloch states we construct projective Wannier functions for the Mn-$3d$ 
orbitals in an energy window from \SI{-5.00}{\electronvolt} to \SI{3.40}{\electronvolt} 
for \ce{BaMn2As2}. Likewise, we choose a window from \SI{-5.50}{\electronvolt} 
to \SI{3.25}{\electronvolt} for \ce{LaMnAsO}. Using such a large energy window for 
the projections results in a much better localization of the Mn-$3d$ Wannier 
functions~\cite{LaFe01,TRIQS/DFTTools,wannier_cubic}, which plays to the strengths 
of the DMFT approximation. 

In DMFT we work with a full rotationally invariant Slater Hamiltonian for the five 
Mn-$3d$ orbitals with a Coulomb interaction $U=F^0$ of \SI{5.0}{\electronvolt} 
and a Hund's coupling $J=(F^2+F^4)/14$ of \SI{0.9}{\electronvolt} ($F^4/F^2 = 0.625$).
We estimate our interaction parameters from the values used in iron pnictide
calculations \cite{ironpnictides_UJ, TRIQS/DFTTools2, haule1},
increasing them slightly to account for the stronger correlations expected in
Mn compounds. It is established that the physics of the nominally half filled
Mn-$3d$ shells is strongly governed by $J$ \cite{georges2013_hund,Janus}.  We find
that our $J$ is consistent with values used in other recent works on manganese
pnictides \cite{LaMnPO01,LaMnPO02, Ba14}. 

Due to the localized nature of the compounds and the substantial electron-electron 
correlations we choose the fully localized limit (FLL) as double counting 
correction~\cite{FLL}. In general the choice of the double counting is less 
crucial in fully charge self-consistent calculations~\cite{TRIQS/DFTTools2}.

The TRIQS/CTHYB solver (v1.3)~\cite{TRIQS/CTHYB}, which is based on continuous-time quantum 
Monte Carlo in the hybridization expansion~\cite{CTQMC,werner2}, is used to obtain the solution 
of the impurity model on the Matsubara axis at an inverse temperature $\beta= \SI{40}{\electronvolt^{-1}}$, 
corresponding to room temperature. We use the stochastic maximum entropy method~\cite{MaxEntBeach} for 
the analytic continuation of the self-energy to the real-frequency axis. In the antiferromagnetic case 
the DFT part is performed without spin polarization; thus the magnetic splitting of the Mn-$3d$ spins is 
purely introduced by DMFT. To describe the desired antiferromagnetic state, the same self-energy is taken 
for both Mn atoms in the unit cell, but with swapped spins. 

We calculate the optical properties within the Kubo formalism, neglecting vertex corrections 
(strongly suppressed in DMFT~\cite{DMFT2,optic_DMFT1}), as implemented in the TRIQS/DFTTools 
package~\cite{TRIQS/DFTTools}. For the optical calculations we increase the number of $k$ points to 
\num{150000} for \ce{BaMn2As2} and \num{100000} for \ce{LaMnAsO} in the full BZ. The frequency-dependent 
optical conductivity is given by
\begin{multline}
	\label{eqn:optcon}
  \sigma^{\alpha\beta}(\Omega)=N_{\sigma}\pi e^2 \hbar \int\!\! d\omega\,
  \Gamma^{\alpha\beta}(\omega + \Omega,\omega)
  \\\cdot
  \frac{f(\omega)-f(\omega + \Omega)}{\Omega},
\end{multline}
with the spin degeneracy $N_\sigma$, the Fermi function $f(\omega)$, and the transport distribution
\begin{multline}
  \label{eqn:Gamma}
  \Gamma^{\alpha\beta}\left(\omega_1,\omega_2\right) =
  \\
  \frac{1}{V}\sum_{\mathbf{k}}
  \mathrm{tr}\left[v^{\alpha}(\mathbf{k})A(\mathbf{k},\omega_1)v^{\beta}(\mathbf{k})A(\mathbf{k},\omega_2)\right],
\end{multline}
where $V$ is the unit cell volume. The summation over $k$ points is properly normalized with respect to their 
weights. The velocities $v^\alpha(\mathbf{k})$, which are proportional to the matrix elements of the momentum 
operator in direction $\alpha = \{x,y,z\}$, are calculated with the \texttt{optic} module of Wien2k~\cite{wien2k_optic}. 
In multiband systems the velocities $v^{\alpha}_{ij}(\mathbf{k})$ and the spectral function $A_{ij}(\mathbf{k},\omega)$ 
are Hermitian matrices in the band indices $i,j$.

To analyze the influence of structural differences, we construct maximally localized Wannier functions and real-space 
Hamiltonians with the help of wien2wannier~\cite{wien2wannier} and Wannier90~\cite{wannier90}.

\section{\label{sec:Resu}Results}
\subsection{\label{subsec:Elec}Electronic Structure}
\begin{figure}[t]
\includegraphics[width=0.97\linewidth]{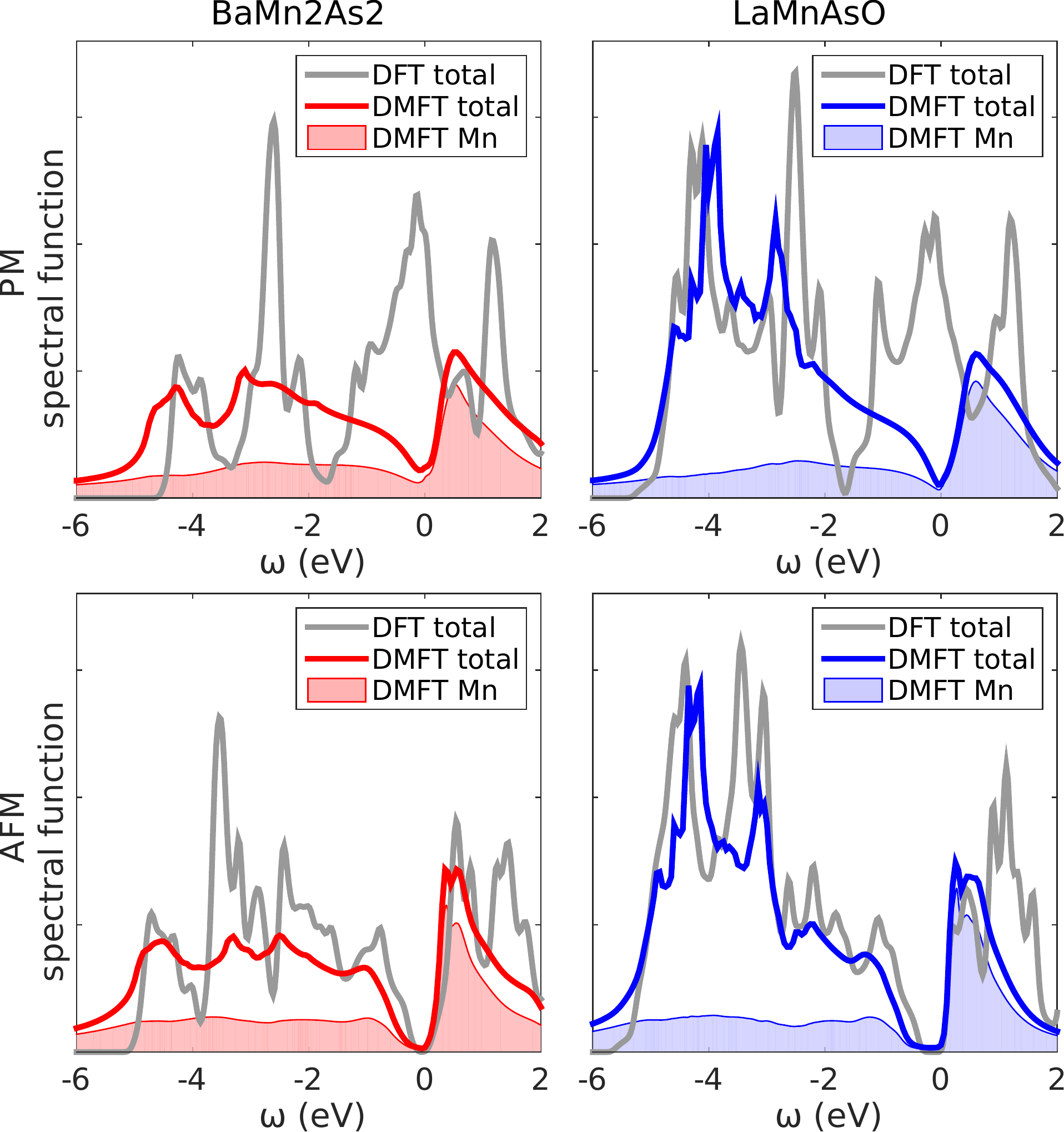}
\caption{\label{fig:DOS}DFT+DMFT paramagnetic (top row) and antiferromagnetic (bottom row) spectral functions of 
\ce{BaMn2As2} (red) and \ce{LaMnAsO} (blue) compared to DFT (gray). The shaded areas correspond to the Mn-projected 
spectral functions. The Fermi level is set to $\omega = \SI{0}{\electronvolt}$.}
\end{figure}

In this section we present our DFT+DMFT results. While we focus on the antiferromagnetic 
(AFM) phase, we also consider the paramagnetic (PM) solution to gain insight into the origin 
of the insulating state (Mott or Slater mechanism).  

We start our discussion with the PM spectral functions in Fig.~\ref{fig:DOS} (top). 
In DFT, the total non-spin-polarized DFT spectral functions (light gray lines) in 
both compounds are clearly metallic, in contrast to DFT+DMFT, where the weight at 
the Fermi level is drastically reduced, nearly opening a gap. The structure
of the correlated Mn-$3d$ spectral function (shaded areas in Fig.~\ref{fig:DOS}) is very 
similar in both compounds and consists of a heavily smeared-out contribution 
below the Fermi energy, which shows strong hybridization with the As-$4p$ bands along 
its full width of about \SI{5.0}{\electronvolt}, and a sharp peak ranging 
from \SI{0.0}{\electronvolt}  to about \SI{2.0}{\electronvolt}. On the other hand, the total spectral 
functions (thick lines in Fig.~\ref{fig:DOS}) differ below \SI{-2.0}{\electronvolt}
due to oxygen states present in \ce{LaMnAsO} but not in \ce{BaMn2As2}.

\begin{figure}[t]
\includegraphics[width=0.97\linewidth]{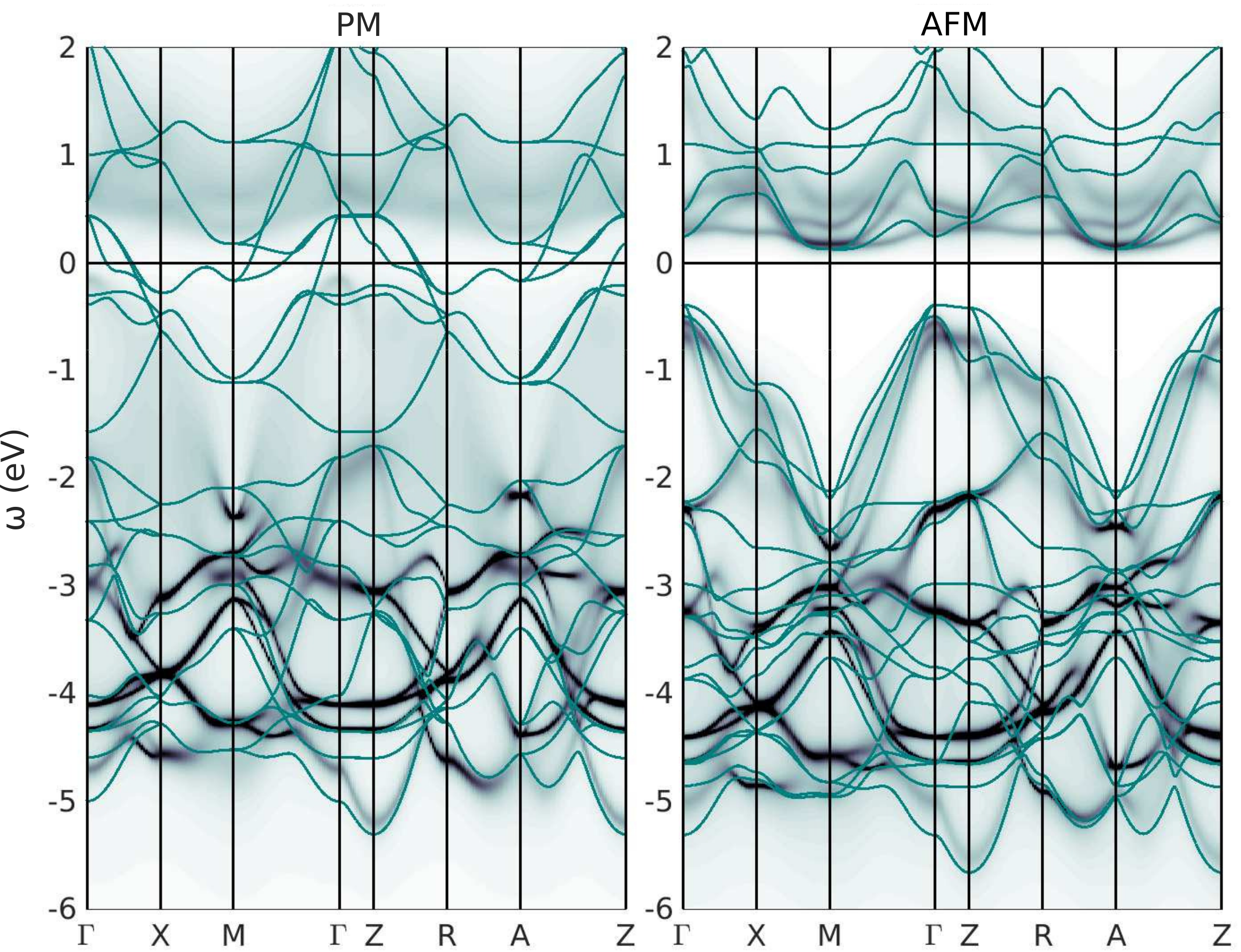}
\caption{\label{fig:BSLa}Spectral function $A(\mathbf{k},\omega)$ for the paramagnetic (left) and 
antiferromagnetic (right) state of \ce{LaMnAsO}. The thin solid lines show the DFT bands, 
while the shading shows the DFT+DMFT spectral weight. The Fermi level is set to 
$\omega = \SI{0}{\electronvolt}$.}
\end{figure}

\begin{figure}[t]
\includegraphics[width=0.97\linewidth]{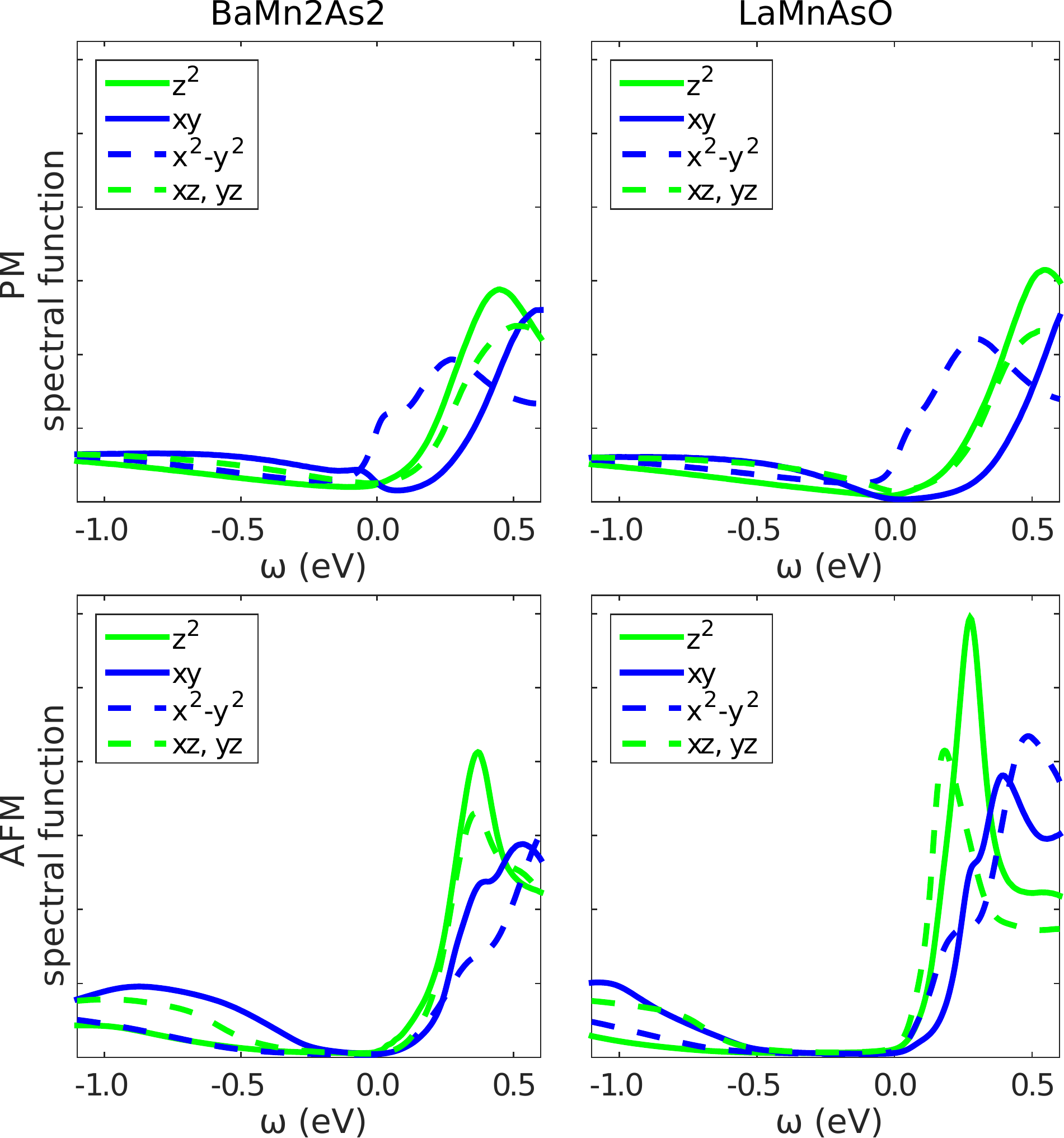}
\caption{\label{fig:DOSMn}DFT+DMFT orbital-resolved paramagnetic (top
  row) and antiferromagnetic (bottom row) spectral functions of the
  correlated manganese atom for \ce{BaMn2As2} (left) and \ce{LaMnAsO}
  (right). The Fermi level is set to $\omega = \SI{0.0}{\electronvolt}$.}
\end{figure}

The strong electron-electron correlations in the half-filled Mn-$3d$ shells 
place both compounds near the Mott metal-insulator transition. This is also seen in 
the $k$-resolved paramagnetic spectral function in Fig.~\ref{fig:BSLa}. At the Fermi 
energy, no remnants of DFT bands are observable, indicating that the spectral weight 
is solely introduced by the imaginary part of the self-energy. This picture is 
supported by the quasiparticle weights, which are below \num{0.13} for all 
orbitals in both compounds. In the orbital-resolved PM spectral functions of the Mn-$3d$ 
shell (Fig.~\ref{fig:DOSMn} top row), we observe orbital-selective behavior: While 
the $x^2-y^2$ orbital is still unequivocally metallic, electronic correlations 
drive the $xz,yz$, and $z^2$ orbitals close to the insulating phase. The $xy$ orbital 
is even more correlated.

To substantiate our claim that the investigated compounds are close to
a Mott transition we perform additional calculations for \ce{LaMnAsO}.
At room temperature and an interaction strength of $U=\SI{6.0}{\electronvolt}$ 
and $J=\SI{1.0}{\electronvolt}$ it remains metallic. However, both lower temperature ($\beta = \SI{100}{\electronvolt^{-1}}$) and stronger interactions ($U=\SI{7.0}{\electronvolt}$ and $J=\SI{1.1}{\electronvolt}$) are independently sufficient to drive the material into the insulating phase.  Hence we conclude that our compounds are indeed close to an insulating phase.

It has been shown before that in comparison to \ce{BaFe2As2} electronic correlations
have a stronger effect in \ce{BaMn2As2}, placing the latter significantly closer 
to the Mott localization picture~\cite{Ba07,BaFe01}. Our paramagnetic DFT+DMFT results
confirm this observation and extend it to \ce{LaMnAsO}, which is also more localized
than its itinerant Fe relative~\cite{LaFe01}.

Conversely, the AFM spectral functions do feature a gap (Fig.~\ref{fig:DOS}
bottom).  DFT predicts small gaps of \SI{0.1}{\electronvolt} for \ce{BaMn2As2} and \SI{0.5}{\electronvolt} 
for \ce{LaMnAsO}, consistent with earlier theoretical results~\cite{Ba03,Ba11,La01,La05}. In DFT+DMFT, 
the gap remains similar in \ce{BaMn2As2} but is somewhat enlarged in \ce{LaMnAsO}, to about \SI{0.6}{\electronvolt}. 
In the case of \ce{BaMn2As2}, the strong incoherence, the finite temperature, and the influence of the analytic 
continuation prohibit the statement of an exact value for the band gap. Nevertheless, the gap is clearly very 
narrow in \ce{BaMn2As2} and of the same order as the DFT result.

Although the increase in the band gaps, due to the DFT+DMFT treatment, is smaller than \SI{0.1}{\electronvolt}, 
electronic correlations lead to important differences in the spectral functions relative to DFT: First, a strong 
renormalization of the bandwidth of the unoccupied Mn states, and second, a substantial smearing of the occupied 
Mn spectral weight (see Figs.~\ref{fig:DOS} and~\ref{fig:BSLa}).

Experiments indicate that \ce{BaMn2As2} has, at least at low temperatures, an indirect band gap of about 
\SI{0.03}{\electronvolt}~\cite{Ba02,Ba01,Ba08,Ba09}. To our knowledge, the only experimental results for the 
\ce{LaMnAsO} gap are \SI{1.1}{\electronvolt} from resistivity measurements of a polycrystalline sample at high 
temperatures~\cite{La01} and \SI{1.4}{\electronvolt} for \ce{LaMnAsO} thin films, deduced from optical absorption 
spectra~\cite{La02}. 

In both materials the fundamental gap is indirect, and the smallest direct gap occurs at the $\Gamma$ point. 
For \ce{BaMn2As2}, this direct gap is about \SI{0.7}{\electronvolt}, close to the recently published value
of \SI{0.8}{\electronvolt}~\cite{Ba14}.  For \ce{LaMnAsO}, we find a direct gap of about \SI{0.8}{\electronvolt} 
(see also Fig.~\ref{fig:BSLa}). Unsurprisingly, the AFM $k$-resolved spectral function and the indirect/direct
gaps of \ce{LaMnAsO} agree well with DFT+DMFT calculations for the closely related manganese pnictide \ce{LaMnPO}~\cite{LaMnPO02}.

Since both compounds share the structure of the \mbox{Mn-As} layers (see Fig.~\ref{fig:crysstruct}), it is natural to 
ask why the band gap of \ce{BaMn2As2} is narrower. The orbital-resolved AFM DFT+DMFT spectral functions projected on 
the Mn-$3d$ states (Fig.~\ref{fig:DOSMn} bottom) reveal that the gap is between the $z^2$ orbital on the unoccupied side 
and the $xy$ orbital on the occupied side in \ce{BaMn2As2}. On the other hand, in \ce{LaMnAsO} the $xy$ gap is considerably 
wider. Therefore, within the assumptions of our calculation, we can attribute the narrower band gap of \ce{BaMn2As2} to the 
different spectral contributions of the $xy$ and the $z^2$ orbitals. Interestingly, the structural difference between the 
investigated compounds mainly impacts those two orbitals, as we will see in the next section.

\subsection{\label{subsec:MLWF}Maximally localized Wannier functions}

\begin{figure}[t]
\includegraphics[width=0.45\linewidth]{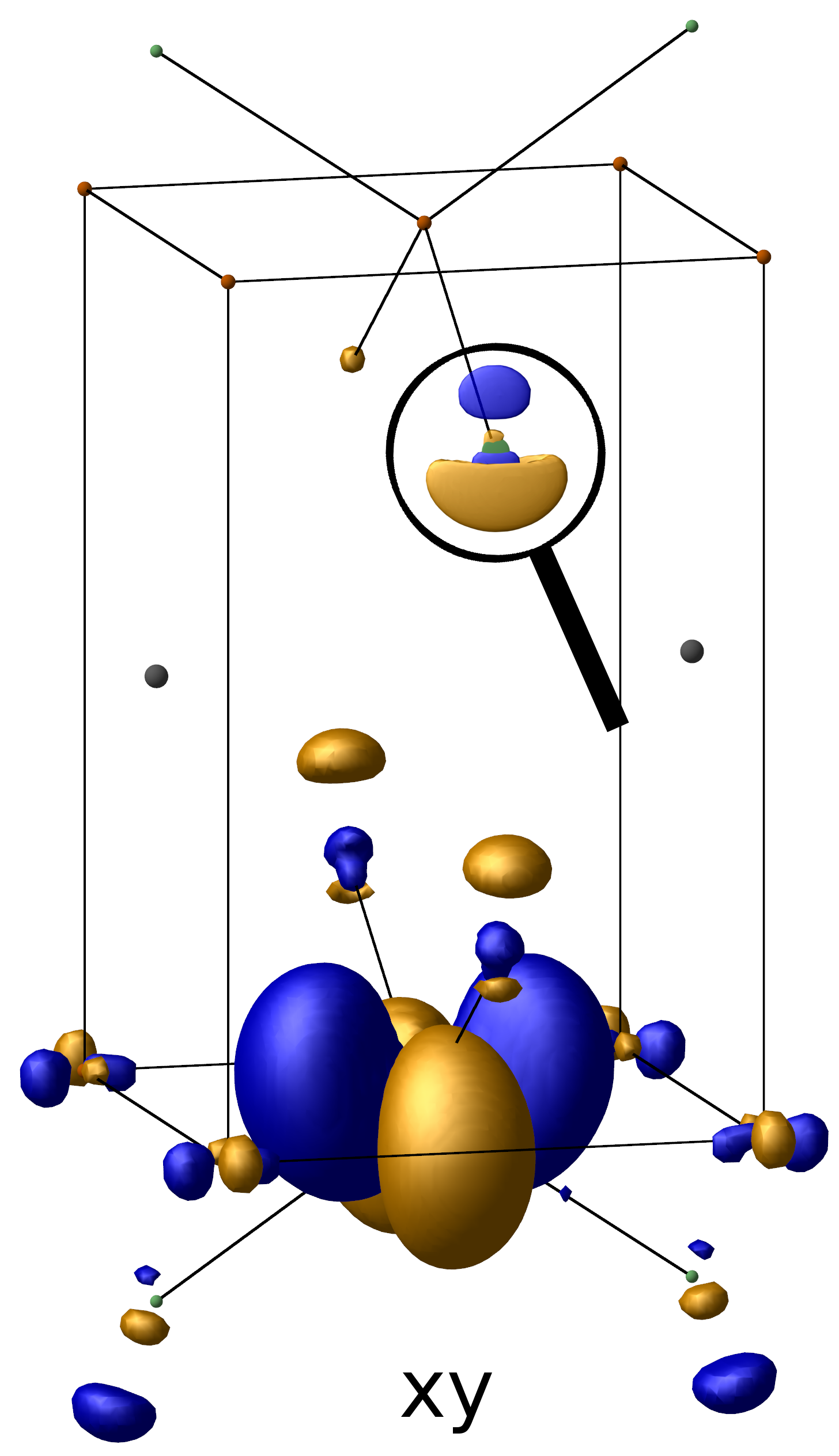}
\includegraphics[width=0.45\linewidth]{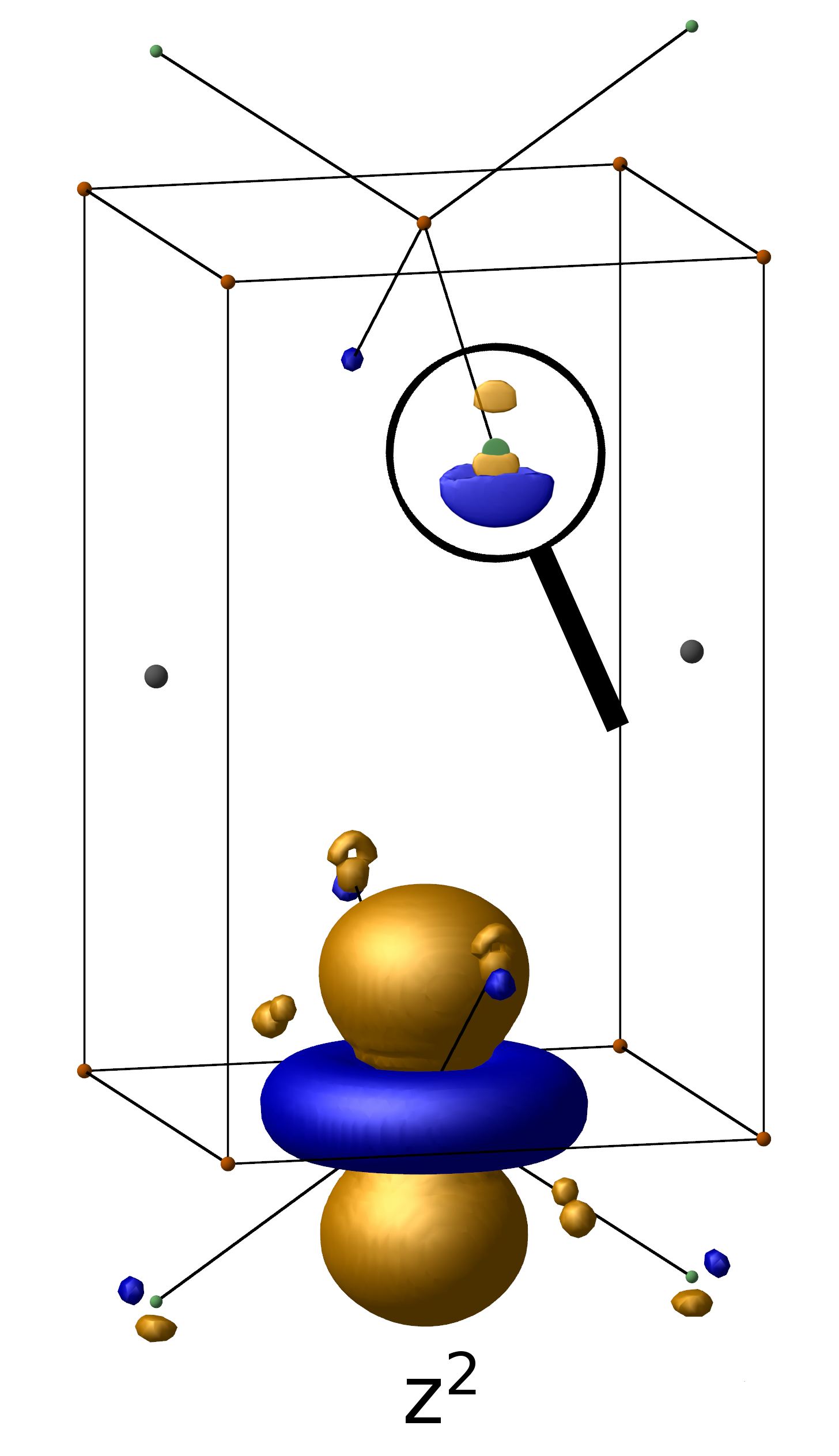}\\
\vspace*{3mm}
\includegraphics[width=0.45\linewidth]{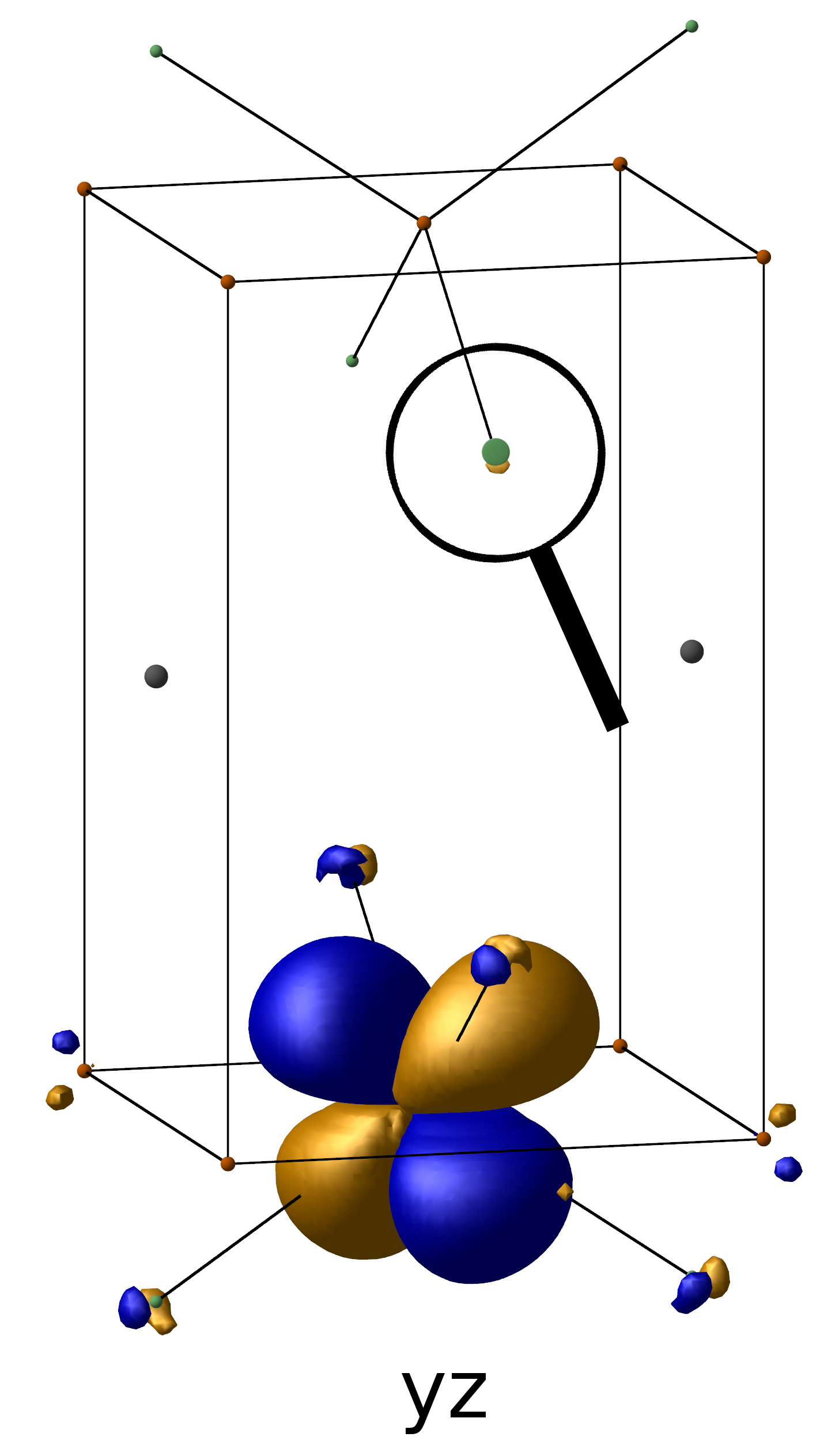}
\includegraphics[width=0.45\linewidth]{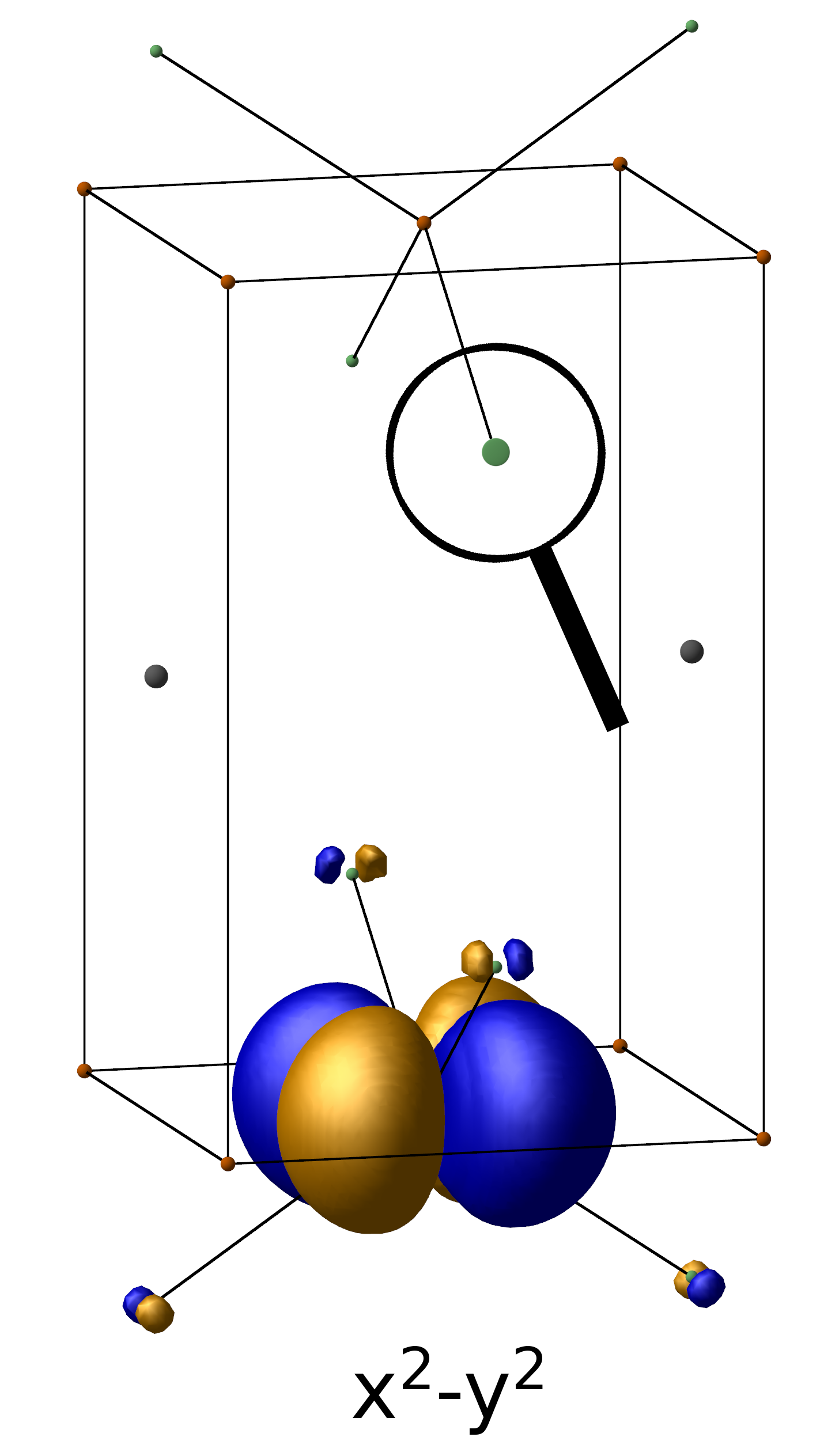}
\caption{\label{fig:BaOrbs} Real-space representation of the maximally localized Wannier orbitals 
for the Mn-$3d$ shell of \ce{BaMn2As2} constructed with wien2wannier~\cite{wien2wannier} and 
Wannier90~\cite{wannier90} and visualized in VESTA~\cite{VESTA}. The $xz$ orbital, which is not 
shown here, is related to the $yz$ orbital by crystal symmetry. The thin lines connect the central 
Mn atoms to the four nearest As atoms. The $xy$ and $z^2$ orbitals have significant weight also on 
the As atoms of the neighboring \mbox{Mn-As} layers. To emphasize this contribution, in the magnifier 
symbols we show it enlarged both by applying a zoom and selecting a smaller isovalue (by a factor of ten). 
By contrast, in \ce{LaMnAsO}, no weight would be seen on the adjacent \mbox{Mn-As} layers at these isovalues.}
\end{figure}

To understand the influence of the structural differences on the Mn-$3d$ orbitals, and on the 
resulting physical properties, we construct an effective real-space Hamiltonian for both compounds.
For the present section, we set aside the projective Wannier functions we use in DFT+DMFT and 
construct ten maximally localized Wannier functions from the non-spin-polarized Mn-$3d$ bands. 
This model has the advantage that it directly provides the effective hopping between Mn atoms, 
including all hopping paths over intermediate atoms (Ba, La, As, O). Thus, it allows us to compare 
the two compounds on the same footing. The maximally localized Wannier functions for the 
Mn-$3d$ orbitals are shown in Fig.~\ref{fig:BaOrbs}.

\begin{figure}[t]
\includegraphics[width=0.97\linewidth]{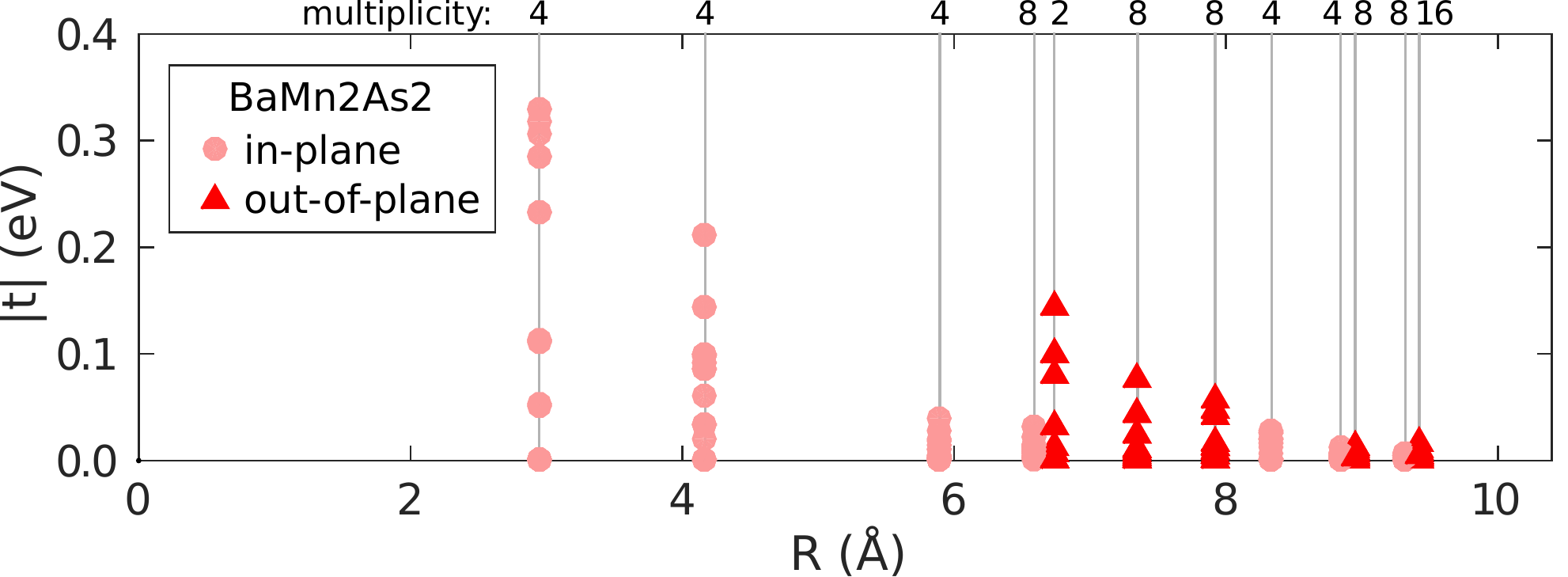}
\includegraphics[width=0.97\linewidth]{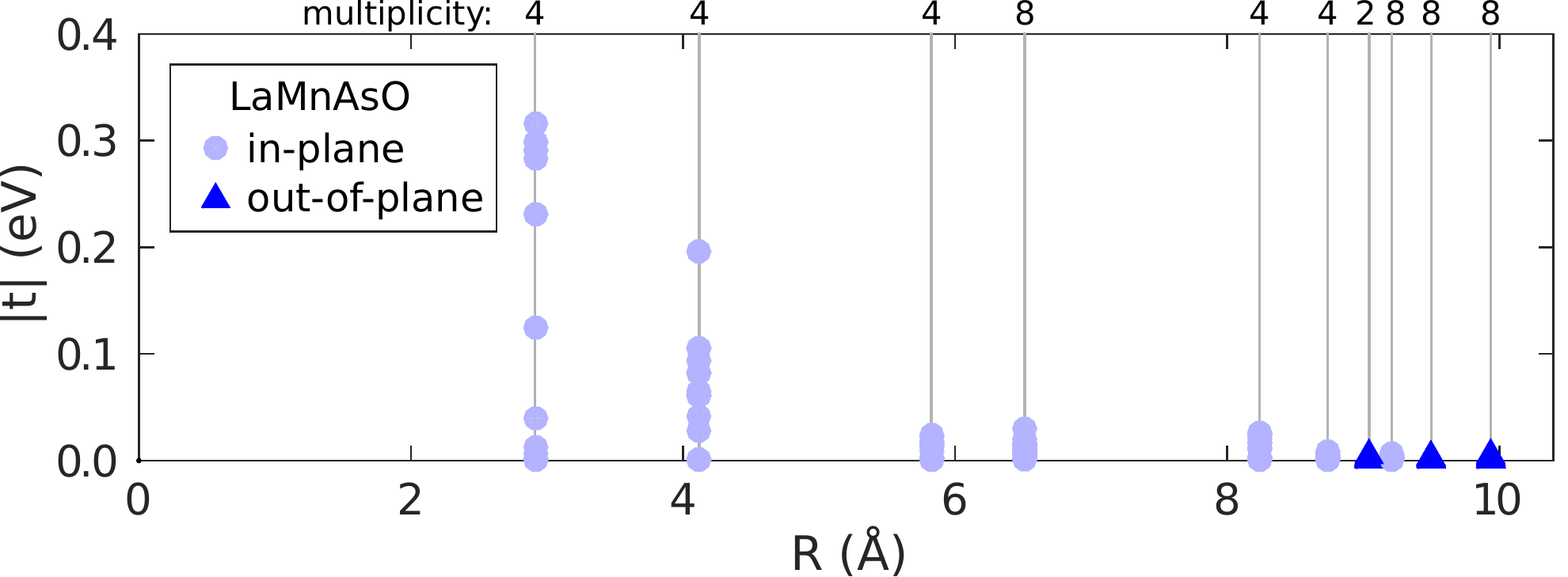}
\caption{\label{fig:Hopping} Real-space Hamiltonian matrix elements $|t|$ for \ce{BaMn2As2} (red) 
and \ce{LaMnAsO} (blue) from a Wannier90 construction of the Mn-$3d$ orbitals. Shown are all hoppings 
between Mn atoms separated by the distance $R$. The given multiplicities correspond to the number of 
neighbors at that distance.}
\end{figure}

The matrix elements of the resulting effective real-space Hamiltonian as a function of 
distance are plotted in Fig.~\ref{fig:Hopping}. For each pair of Mn atoms all \num{25} 
matrix elements between their five Mn-$3d$ orbitals are shown. Considering the in-plane hopping 
first (circles), both materials are described by a very similar Hamiltonian, which is expected 
due to the shared structure of the \mbox{Mn-As} layers and the comparable Mn-Mn distance within 
those layers. Turning to the interlayer hoppings (triangles), a completely different picture 
emerges. The Mn atoms in \ce{BaMn2As2} couple substantially to their respective neighbors on 
adjacent planes, in contrast to \ce{LaMnAsO}, where the interlayer coupling is on average more 
than \num{25} times lower and not visible on the shown scale. From this it follows that \ce{LaMnAsO} 
is built up by quasi-two-dimensional \mbox{Mn-As} layers coupled only very weakly to each other, 
whereas \ce{BaMn2As2} shows much stronger interlayer coupling. 

It bears mentioning that in \ce{BaMn2As2} the interlayer hoppings follow a different 
decay than the in-plane hoppings with distance, as visible in the much stronger coupling 
for similar atomic distances. The responsible hoppings for the interlayer coupling can 
be nearly exclusively attributed to the $xy$ and $z^2$ orbitals, as the coupling of 
the $xz$, $yz$ orbitals is already below \SI{0.05}{\electronvolt}. The $x^2-y^2$ orbitals 
practically do not contribute to the interlayer coupling. The largest hopping strength is 
found between the $xy$ orbitals, followed by $xy$ to $z^2$ and $z^2$ to $z^2$. 

The difference in the interlayer coupling of the \mbox{Mn-As} layers can be traced back 
to multiple factors. First, the distance between the \mbox{Mn-As} layers is much shorter 
in \ce{BaMn2As2} (\SI{6.73}{\angstrom} versus \SI{9.04}{\angstrom}), where they are 
separated only by the rather narrow Ba layer in contrast to the thicker La-O layer in 
\ce{LaMnAsO} (see Fig.~\ref{fig:crysstruct}). The fact that the in-plane coupling on 
distances comparable to the interlayer distance is substantially smaller than the out-of-plane 
coupling indicates that the spatial distance between the layers is not enough to fully 
explain the enhanced out-of-plane coupling. The second important difference is the stacking
inversion of the \mbox{Mn-As} layers in \ce{BaMn2As2}. In \ce{LaMnAsO} the As atoms do 
not sit directly above each other as they do in \ce{BaMn2As2}, where the small As-As 
distance leads to a simple hopping path via the As atoms. In the maximally localized 
Wannier functions, this can be seen in the electronic weight of the $xy$ and $z^2$ 
orbitals on the As atoms in the adjacent layer (Fig.~\ref{fig:BaOrbs} magnifier symbol).

\subsection{\label{subsec:Neel}N\'eel temperature}
The highest N\'eel temperature $T_{\text{N}}$ in a class of compounds is usually found close to the 
Mott transition. This was first shown for the single-band Hubbard model on the Bethe lattice~\cite{Neel05} 
and the same argument was recently found to be valid for the $4d$ perovskite \ce{SrTcO3}~\cite{Neel04} with its 
exceptionally high transition temperature of \SI{1000}{\kelvin}. When starting from an itinerant picture, 
for a model with bandwidth $W$, where the interactions of order $U$ are treated on a mean-field level,
the transition temperature scales as $T_{\text{N}} \sim \exp(-W/U)$. On the fully localized side, the 
adequate picture is the Heisenberg model, where the scaling is $T_{\text{N}} \sim W^2/U$. Between these 
two extreme cases we can expect a crossover around $U \approx W$, which coincides with the crossover 
from an itinerant to a localized system. These qualitative considerations identify the Mott transition as a 
hotspot for magnetism, and hence, materials in this critical region are prone to higher transition temperatures.  

\begin{figure}[t]
\includegraphics[width=0.97\linewidth]{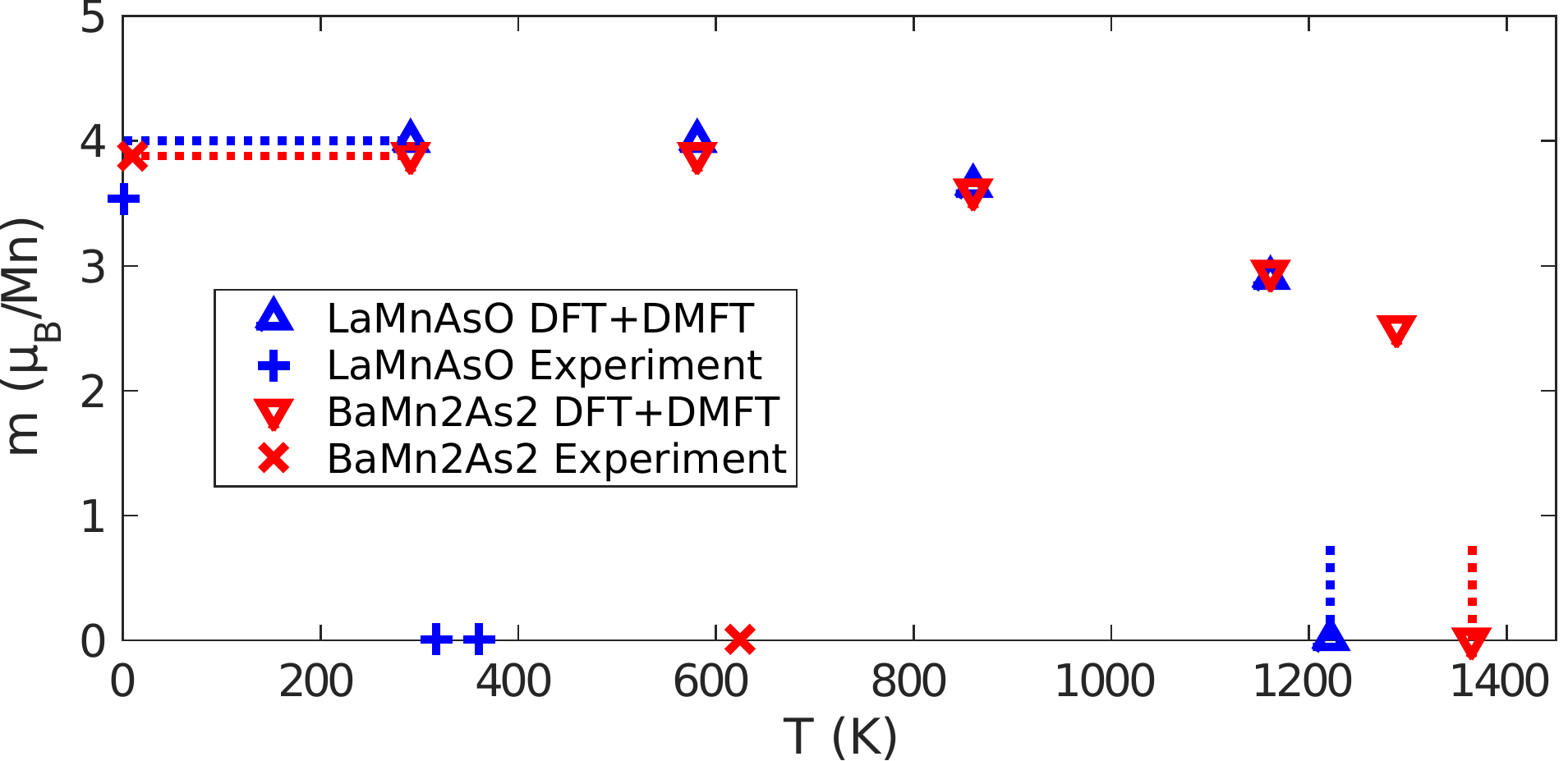}
\caption{\label{fig:mvsT}Magnetic moment $m$ versus temperature $T$ from fully charge self-consistent DFT+DMFT for 
\ce{BaMn2As2} (red triangles) and \ce{LaMnAsO} (blue triangles). Experimental points are taken from 
Refs.~\onlinecite{La10,La06,La06_sup} for \ce{LaMnAsO} (blue crosses) and from Ref.~\onlinecite{Ba06} for
\ce{BaMn2As2} (red crosses). The horizontal dotted lines mark the saturated magnetic moments and the vertical dotted 
lines the approximate N\'eel temperatures.} 
\end{figure}

From the paramagnetic DFT+DMFT spectral function and the quasiparticle weights, we have seen that \ce{BaMn2As2} 
and \ce{LaMnAsO} are close to a Mott transition, and their experimental N\'eel temperatures are indeed high, with 
reports of  \SI{317}{\kelvin} to \SI{360}{\kelvin} in \ce{LaMnAsO}~\cite{La06, La10} and even \SI{625}{\kelvin} in
\ce{BaMn2As2}~\cite{Ba06}. In Fig.~\ref{fig:mvsT} we present our DFT+DMFT results for the ordered moment as a 
function of temperature. For \ce{BaMn2As2} we find an ordering temperature of around \SI{1350}{\kelvin}. Taking 
into account the overestimation of the transition temperature expected due to the mean-field character of 
DMFT~\cite{Neel04}, which is roughly a factor of two~\cite{Neel06, Neel04}, the prediction of the N\'eel temperature 
in our calculation is in reasonable agreement with the experimental value. Also, the ordered moment at low temperatures 
agrees well with the experimental result of \SI{3.9}{\muB\per Mn}~\cite{Ba06}.

The situation is different for \ce{LaMnAsO}. There, the experimental ordering temperature is a factor of two smaller 
than for \ce{BaMn2As2}. However, the DFT+DMFT result is smaller only by about \SI{150}{\kelvin}. This observation is 
very similar to recent studies in technetium oxides. In the cubic case (\ce{SrTcO3}), the local DMFT approximation 
works well~\cite{Neel04}, but for the layered counterpart \ce{Sr2TcO4} it overestimates the ordering temperature 
substantially~\cite{sr2tco4}. The reason is that in quasi-two-dimensional systems, as is the case for \ce{Sr2TcO4} and
also for \ce{LaMnAsO}, spatial fluctuations become important. They in turn decrease the ordering temperature significantly. 
In the same way, DFT+DMFT yields a saturated magnetic moment of \SI{4.0}{\muB\per Mn} in contrast to the measured 
\SI{3.6}{\muB\per Mn}~\cite{La06,La06_sup}.
 
From another point of view, it is well known that the strength of the interlayer coupling is a crucial factor influencing 
the magnetic properties of layered materials~\cite{Neel03}, for instance in the copper oxides~\cite{Neel01,Neel02}. In such 
compounds, the crossover from a three-dimensional to a layered system leads to a suppression of the N\'eel temperature, as 
a function of the interlayer exchange coupling $J_\perp$. As we have observed above, the strong decrease of dimensionality 
in \ce{LaMnAsO} is confirmed by a reduction of the hopping in the $z$ direction (by a factor larger than \num{25}). This will also be
reflected in an effective $J_\perp$. Band theoretical estimates and experiments suggest that $J_\perp/J_\parallel < \num{0.015}$ 
in \ce{LaMnAsO}~\cite{La05} and $J_\perp/J_\parallel \approx 0.1$ in \ce{BaMn2As2}~\cite{Ba10}, with $J_\parallel$ being the 
in-plane nearest neighbor exchange coupling. This supports the conclusion that the reduced N\'eel temperature in \ce{LaMnAsO} can 
be attributed to the lower dimensionality of the system. 

\subsection{\label{subsec:Optic}Optical Properties}
Finally, we turn to the optical conductivity $\sigma^{\alpha\beta}(\Omega)$, Eq.~(\ref{eqn:optcon}), which will not only exemplify 
the dimensional differences of the investigated compounds but also allow a comparison with experimental observations.

\begin{figure}[t]
\includegraphics[width=0.97\linewidth]{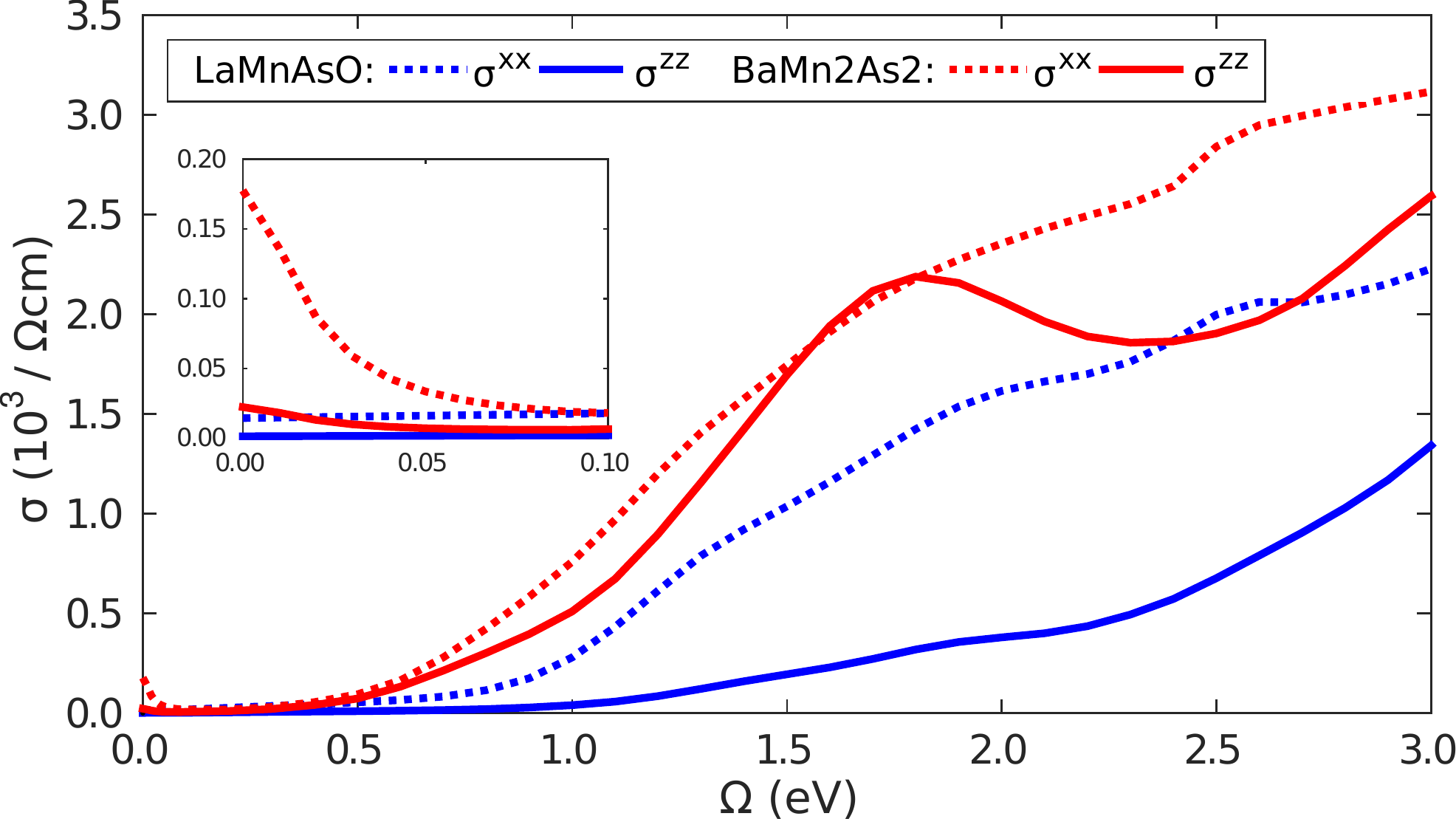}
\caption{\label{fig:OPTICdir} Optical conductivity tensor components $\sigma^{zz}(\Omega)$ (solid lines) and $\sigma^{xx}(\Omega)$ (dotted lines) 
of \ce{BaMn2As2} (red) and \ce{LaMnAsO} (blue) from fully charge self-consistent DFT+DMFT including uncorrelated bands outside the projective 
window. The inset shows the low frequency region of the optical conductivity in the same units as the main panel.}
\end{figure}

Starting with \ce{BaMn2As2} (Fig.~\ref{fig:OPTICdir} red lines), we observe only a weak anisotropy in the optical conductivity. 
Besides the depression around $\Omega \approx \SI{2.5}{\electronvolt}$, there is not much of a difference between the in-plane and 
out-of-plane contributions. This illustrates the more isotropic nature of \ce{BaMn2As2}. A small Drude peak present in the in-plane 
component is in accordance with the observation of a weak metallic behavior at room temperature in Refs.~\onlinecite{Ba01,Ba02}. On 
the contrary, there is no Drude peak in the out-of-plane component, indicating that \ce{BaMn2As2} is insulating along the $z$ direction, 
in agreement with the optical experiments of Ref.~\onlinecite{Ba01}. The optical conductivity of \ce{LaMnAsO} (Fig.~\ref{fig:OPTICdir} 
blue lines) in the $x$ direction shows a similar trend, but is reduced by about \num{1/3} in comparison to \ce{BaMn2As2}. Since $\sigma^{zz}(\Omega)$ 
is strongly suppressed in \ce{LaMnAsO}, the total optical conductivity becomes largely dominated by the in-plane contribution.
 
The similar in-plane conductivity of both compounds originates from the common \mbox{Mn-As} layer structure, though the effect of 
the structural differences becomes apparent in the distinct optical properties in the $z$ direction.  We emphasize that the dimensional 
difference is also visible in the DC conductivity ($\Omega \rightarrow 0$). The ratio $\sigma_{xx}(0)/\sigma_{zz}(0)$, an indicator 
for the anisotropy, is \num{17} in \ce{LaMnAsO} but only \num{7} in \ce{BaMn2As2}. 

\begin{figure}[t]
\includegraphics[width=0.97\linewidth]{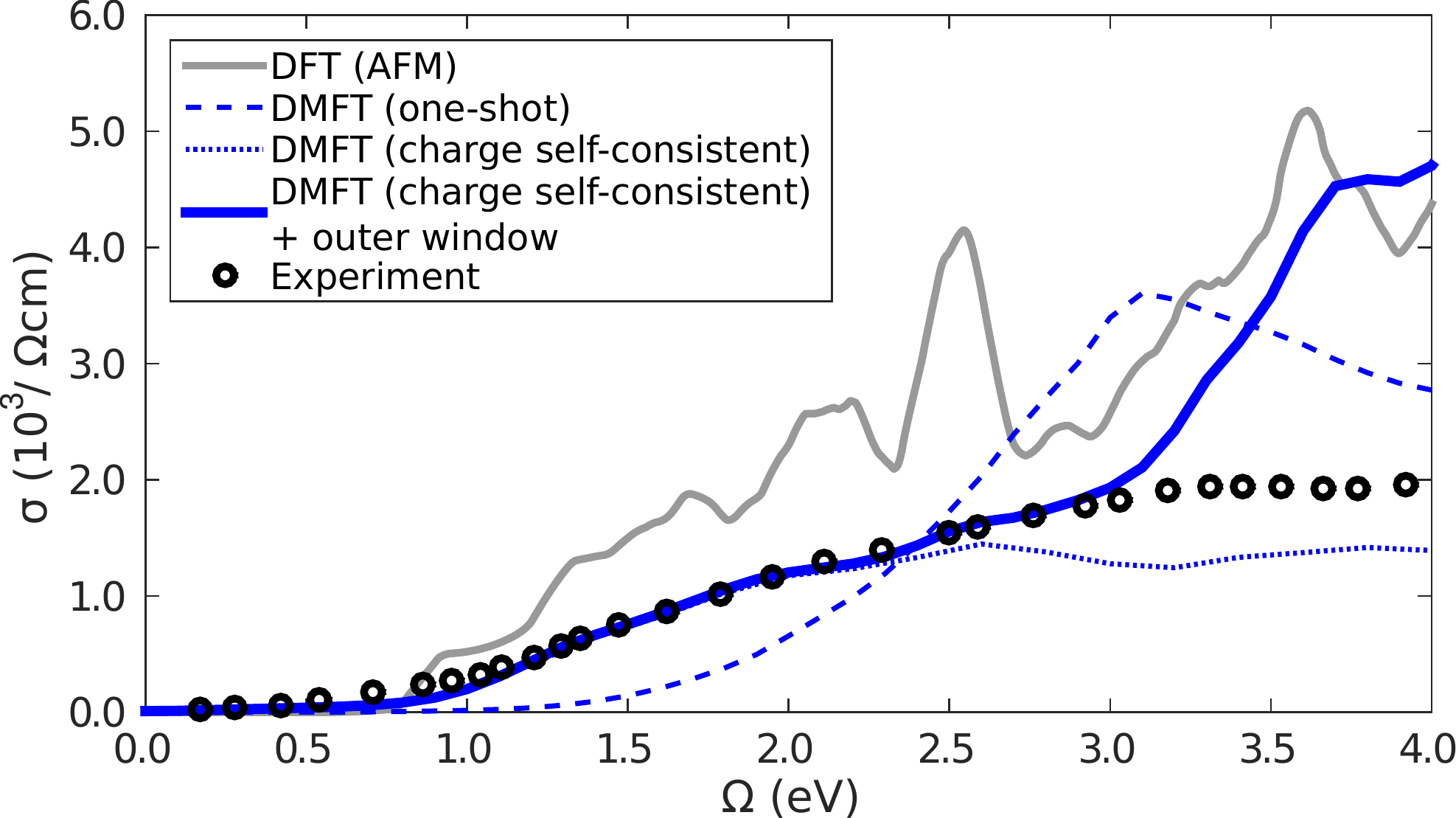}
\caption{\label{fig:OPTICLa} Optical conductivity of \ce{LaMnAsO} calculated with DFT (solid gray line), fully charge self-consistent DFT+DMFT in 
the correlated window (dotted blue line) and including uncorrelated bands (solid blue line) as well as one-shot DFT+DMFT (dashed blue line); 
compared to experimental data (black circles) from Ref.~\onlinecite{La01}. Above \SI{3}{\electronvolt} DFT+DMFT (including the outer window) starts to deviate from the experimental data due to the onset of the La-$4f$ bands, which are placed much too low in energy by DFT~\cite{FLL}.}
\end{figure}

For \ce{LaMnAsO} the optical conductivity was measured using ellipsometry for a polycrystalline sample~\cite{La01}. To compare the 
experimental results to our calculations, we average our theoretical results over all Cartesian directions to obtain a 
``polycrystalline conductivity'' (Fig.~\ref{fig:OPTICLa}). In general, the DFT result (solid gray line) follows the trend of the 
experimental data (black circles), but it severely overestimates the optical conductivity, at some points by more than a factor of 
two. Similarly, a one-shot DFT+DMFT calculation (dashed blue line), where DMFT is converged without updating the charge density, 
cannot explain the experimental data. Here, the static spin splitting leads to a larger gap, which is clearly visible in the suppressed 
optical conductivity below \SI{1.5}{\electronvolt}. Additionally, the unoccupied Mn states are less correlated due to smaller 
electron-electron scattering as compared to full charge self-consistency. This results in a different distribution of the optical weight. 

On the other hand, in the fully charge self-consistent treatment, the optical results are not only influenced by the modified spectral 
function but also by the altered velocities of the updated Kohn-Sham bands. Indeed, the fully charge self-consistent DFT+DMFT calculations 
correctly reproduce the experimental result over a wide range of frequencies (dotted blue line). 

The upper limit of our projective energy window is at \SI{3.25}{\electronvolt}, but the chemical potential ($\mu=\SI{1.6}{\electronvolt}$) effectively shifts this level down
to \SI{1.65}{\electronvolt}. If we also consider that the Fermi energy is close to the unoccupied states, we see that there can be transitions at 
$\Omega \approx \SI{2}{\electronvolt}$, which are not captured with our window. Of course this effect sets in very slowly as there are still many other transitions 
possible at $\Omega \approx \SI{2}{\electronvolt}$. To this end, we extend the trace in Eq.~(\ref{eqn:Gamma}); now, the spectral function is a matrix built by a 
block $A_{ij}$ for the correlated bands, as well as blocks for uncorrelated bands below and above the correlated subspace, $A_{\nu\nu'}$.
Note that $A_{\nu\nu'} \sim \delta_{\nu\nu'}$ is a noninteracting DFT spectral function. The resulting optical conductivity yields excellent
agreement with the experimental data up to about \SI{3}{\electronvolt} (solid blue line). The strong increase above \SI{3}{\electronvolt} is caused by the onset 
of the La-$4f$ bands, which are known to be placed much too low in energy by DFT~\cite{FLL}. 

The remarkable agreement with experimental data underlines the importance of the fully charge self-consistent approach and suggests that our 
choice of the parameters $U$ and $J$ is appropriate. Furthermore, the experimental and theoretical results indicate that the direct band gap of 
bulk \ce{LaMnAsO} may be well below the \SI{1.4}{\electronvolt} obtained from thin-film measurements~\cite{La02}. 

\section{\label{sec:Conc}Conclusion}
We investigated the manganese pnictides \ce{BaMn2As2} and \ce{LaMnAsO} in their paramagnetic and antiferromagnetic phases. These manganates 
represent two points along the dimensional crossover: While \ce{BaMn2As2} is quite isotropic with comparable couplings within and 
between the \mbox{Mn-As} layers, \ce{LaMnAsO} is effectively two dimensional with only a weak residual interlayer coupling. 
This difference, which is already visible in the crystal structure, is substantiated, and its origins are accounted for by the 
maximally localized Wannier functions for the Mn-$3d$ bands and their hopping amplitudes. We demonstrated that differences in physical 
properties such as the N\'eel temperature and DC as well as optical conductivity can be traced back to a large extent to the difference 
in effective dimensionality.

Our fully charge self-consistent DFT+DMFT calculations yield excellent agreement with the experimental optical conductivity. 
Our confidence in the applicability of the method to our compounds thus confirmed, we established that both materials are
near the Mott metal-insulator transition, which helps explain their high N\'eel temperatures.

Our results constitute an important example where charge self-consistent DFT+DMFT is demonstrably superior to the one-shot 
approximation.

\section{\label{sec:Ackn}Acknowledgments}
M.~Zingl, E.~Assmann, and M.~Aichhorn acknowledge financial support from the Austrian Science Fund (Y746, P26220, F04103) as well 
as NAWI Graz and great hospitality at Coll\`ege de France and \'Ecole Polytechnique. P.~Seth acknowledges support from ERC Grant 
No. 278472-MottMetals. I.~Krivenko acknowledges support from Deutsche Forschungsgemeinschaft via Project SFB 668-A3. The authors want 
to thank E.~Schachinger, G.~J.~Kraberger, and R.~Triebl for fruitful discussions. The computational results presented have been achieved 
using the Vienna Scientific Cluster (VSC) and the dcluster of Graz University of Technology.

\bibliography{literatur.bib}

\end{document}